\documentclass[prb,amsmath,superscriptaddress,citeautoscript,twocolumn,
showkeys,showpacs,floatfix]{revtex4}
\usepackage{bm}
\usepackage{amssymb}
\usepackage{graphicx}

\newcommand{\beq}{\begin{equation}}
\newcommand{\eeq}{\end{equation}}

\begin{document}

\title{The conductance of  superconducting-normal hybrid structures}

\author{O. Entin-Wohlman}
\email{oraentin@bgu.ac.il}


 \affiliation{Department of Physics, Ben
Gurion University, Beer Sheva 84105, Israel}

\affiliation{Albert Einstein Minerva Center for Theoretical
Physics, Weizmann Institute of Science, Rehovot 76100, Israel}

\author{Y. Imry}

\affiliation{Department of Condensed Matter Physics,  Weizmann
Institute of Science, Rehovot 76100, Israel}

\author{A. Aharony}

\altaffiliation{Also at Tel Aviv University,
Tel Aviv 69978, Israel}

\affiliation{Department of Physics, Ben Gurion University, Beer
Sheva 84105, Israel}

\date{\today}

\begin{abstract}
The dc conductance of normal-superconducting hybrid structures is discussed.
It is shown that since the Bogoliubov-DeGennes (BDG) equation does not
conserve charge, its application to create a Landauer-type approach for the
conductance of the NSN system is problematic. We `mend' this deficiency by
calculating the conductance from the Kubo formula for a ring configuration
(for this geometry the solutions of the BDG equation conserve charge). We
show that the presence of a superconductor segment within an otherwise normal
metal may {\em reduce} the overall conductance of the composite structure.
This reduction  enhances the tendency  of the NS composite to become
insulating.

\end{abstract}



\maketitle

\section{Introduction}

\label{INTROD}

In a seminal paper published a long time ago, Blonder, Tinkham,
and Klapwijk  \cite{BTK} (BTK)  calculated the conductance of a
normal (N)--superconducting (S) interface as a function of the
interface transparency. In particular they showed that at zero
temperature ($T=0$) and for electrons at the Fermi energy that
conductance is given by
\begin{align}
G^{}_{\rm NS}=\frac{2e^{2}}{h}\Bigl
(1-|S^{ee}_{}|^{2}+|S^{he}_{}|^{2}\Bigr )\ ,\label{BTKFO}
\end{align}
where $|S_{}^{ee}|^{2}$ is the ``normal" reflection coefficient of the
interface, while $|S^{he}_{}|^{2}$ is the reflection coefficient for the
Andreev processes, \cite{ANDREEV} which reflect electron-like excitations as
hole-like ones. For simplicity, both the normal and the superconductor
regions were taken to be free of any impurity scattering (except at the
interface). Then, when the interface is perfectly transparent, i.e.,
$|S_{}^{ee}|^{2}=0$ and $|S^{he}_{}|^{2}=1$, the value of $G^{}_{\rm NS}$ is
twice that of the ``normal" quantum limit of the conductance. However, when a
large enough barrier exists at the interface between the superconducting and
the normal regions, $G^{}_{\rm NS}$ becomes much smaller than the conductance
obtained when the superconductor is made normal.  This is the simplest
example where superconductivity in a part of a system {\it reduces} its
overall conductance. (The more complicated many-channel, disordered case
is not addressed in this paper.) In this paper we  consider the more
subtle NSN combination, and demonstrate a similar effect: with a large enough
barrier at even one of the NS interfaces, the appearance of superconductivity
in the S region reduces the overall conductance.

Some of the motivation for the present work comes from our wish to understand
why the mixed NS bulk composite structure is often insulating at  $T = 0$ and
the superconducting phase of a thin film goes over (with increasing disorder
or decreasing film thickness) directly into the insulating rather than to the
normal-conducting phase \cite{EXP,SHAHAR}. It is interesting that often the
activation energy for the conductance of the insulating phase is given by the
superconductor gap of the superconducting component \cite{SHAHAR}. Thus, the
superconducting component plays the role of an additional  barrier between
the normal segments! The charge-vortex duality \cite{FISHER} for a system of
charged bosons of course explains the insulating phase as dual to the
superconducting one. Vortex localization yields zero resistance and charge
localization yields zero conductance at $T=0$. Our purpose is to provide a
heuristic understanding of how the charge localization is established.
Reducing the small-scale conductance of the system pushes it towards the
insulating state.  The NSN system is the simplest microscopic element of
the NS network. We find that it already presents nontrivial theoretical
questions having to do with a seeming deficiency of the Bogoliubov-DeGennes
(BDG) formulation.

Blonder {\it et al.} \cite{BTK} employed  the BDG
equation for the quasiparticle excitations in the superconducting region,
assuming that the energy gap $\Delta$ which vanishes in the normal part does
not vary spatially in the superconductor. Since in most situations the
superconducting coherence length $\xi$ is much larger than the Fermi
wavelength and much smaller than the length of the S-region, this assumption
seems quite harmless. Consequently, it is widely accepted that the use of the
BDG equation, without attempting to compute the (complex,  in general)
superconducting order-parameter self-consistently, is valid for many  hybrid
structures.

A particularly important issue emphasized by BTK concerns the conversion of
the normal current into a super-current at the NS interface, ensuring
charge-current conservation over the entire structure (see also Ref.
~\onlinecite{KUMEL}). Blonder {\it et al.} \cite{BTK}  showed that the normal
charge-current (which is distinct from the quasiparticle current) entering
the superconductor decays, but concomitantly a super-current grows up
gradually, until very far inside S the normal current disappears completely,
and the entire charge is carried away by  Cooper pairs. One might wonder what
happens to this scenario  when the superconductor has a finite width and is
not infinite as in the BTK case. It turns out that within the BDG formulation
the super-current in the S region does not properly convert back to normal current
in the second N region. We  are not aware of a way to correct for this deficiency
(leaving aside the possibility mentioned above
to compute the order-parameter self-consistently).  In
this paper we will circumvent this problem by using a particular geometry.

Not surprisingly, transport through hybrid NS structures  has been addressed
before, even prior \cite{DEMLER} to the publication of the paper by BTK.
Indeed, as Eq. (\ref{BTKFO}) bears a strong resemblance to the Landauer
formula  \cite{LANDAUER} for coherent transport, several modern treatments
(representative examples are to be found in Refs.
~\onlinecite{JAP,LAMBERT,ANANTRAM,MELIN}) employ scattering theory within the
Landauer picture  in an attempt to extend the BTK result (\ref{BTKFO}) to
more than a single NS interface. The scattering formalism is also used to
study the current fluctuations in mesoscopic systems with Andreev
reflections. \cite{ANANTRAM,MELIN} However, there is a vicious caveat in this
approach: the BDG equation, while conserving the number of quasiparticles,
does not conserve charge. The BDG formulation follows, in this respect, the
similar deficiency of the BCS approach upon which the BDG equation is based.
This does not cause any harm when the formulation is used for a bulk
superconductor, as had been done by BTK (see also Ref.
~\onlinecite{BEENAKKER}), or when the system has the shape of a closed ring,
where the boundary conditions enforce current conservation. \cite{KB}
However, when the size of the superconducting segment is finite, the
non-conservation of the charge leads to current {\it vs.}
chemical-potential-difference relations (within the linear response regime)
that depend on the chemical potential of the superconductor, as opposed to
the situation in the normal phase. \cite{JAP,LAMBERT} Indeed, the
conductances of various hybrid structures have been determined within
scattering theory (when the superconducting is ``floating") by fixing the
chemical potential on S such that current is conserved. \cite{JAP,LAMBERT}
One would have assumed na\"{\i}vely that this procedure cures the problem
mentioned above. It turns out, however, that it does not: as discussed by
Anantram and Datta \cite{ANANTRAM} (see also Ref. ~\onlinecite{MELIN}),
applying the very same procedure to the calculation of the current
fluctuations (the power spectrum of the noise) violates the Johnson-Nyquist
relationships.  We show in Sec. \ref{DISCA}  that this `floating-potential' approach
also does not produce the doubling of the conductance [see Eq. (\ref{BTKFO}) above]
when the NS interface becomes transparent. It approaches, however, the conductance found by the Kubo formula in the limit where the barrier is strong and its transparency is low.

\begin{widetext}

\begin{figure}[h]
\includegraphics[width=7.cm]{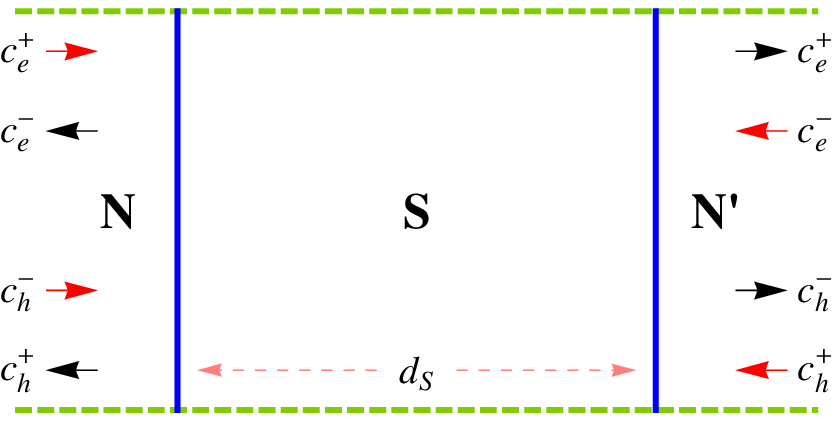}
\hspace{2cm}
\includegraphics[width=6.cm]{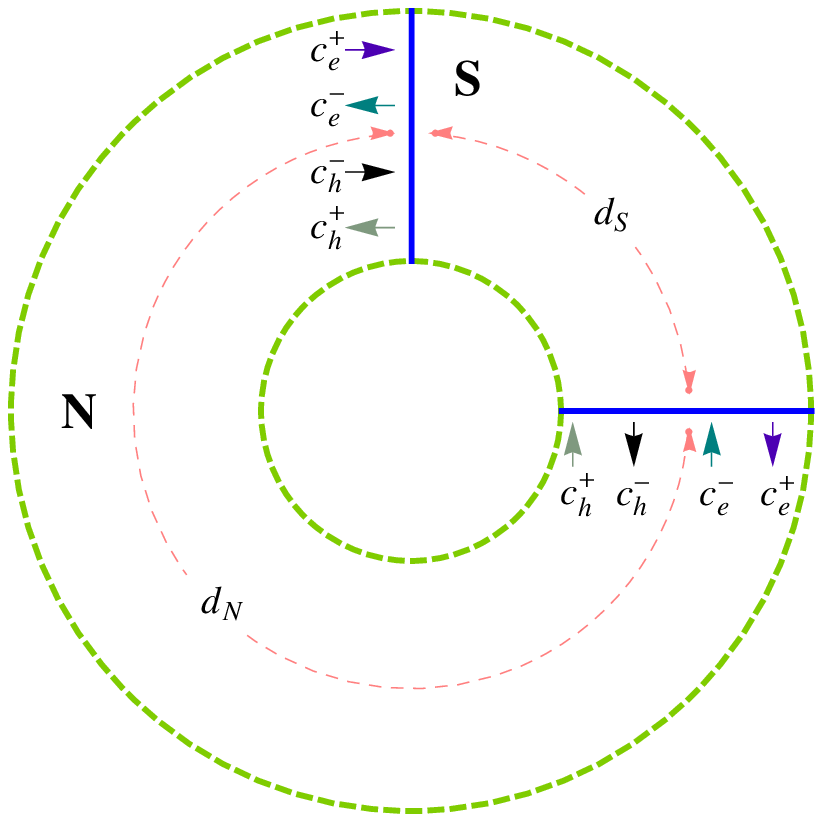}
\caption{Left panel: the `open' NSN structure, where the
perpendicular lines represent potential barriers (the explicit
model treated in Sec. \ref{MAIN} assumes that the right interface
is clean). Right panel: the corresponding ring NSN system. The
incoming and outgoing amplitudes of the electron-like and the
hole-like waves on both sides of the superconductor are marked by
arrows.} \label{SYS}
\end{figure}

\end{widetext}

It follows that in order to use the BDG equation for the calculation of  the
conductance of a mesoscopic system containing a finite-length superconducting
segment,  one needs charge-conserving solutions of that equation. Such
solutions arise naturally when the NSN structure (the left panel in Fig.
\ref{SYS}) is closed to form a ``ring" (the right panel in Fig. \ref{SYS}).
One can use those solutions in the Kubo formula for the conductivity of a
large system.  This is the approach adopted in this paper.

We address the simplest problem of a one-dimensional NSN structure at zero
temperature (we do not discuss here the complications arising  in the
multi-channel case including disorder). After discussing the conductance
between the two N
segments (left panel in Fig. \ref{SYS}),  and dealing with the nontrivial problem arising from the
above-mentioned deficiency of the BDG picture, we find that the NSN
conductance can indeed {\em decrease} when, for example, the superconducting
component becomes longer. We are not able to  treat here the behavior of the
two-dimensional  or the three-dimensional  NSN arrays. Suffice it to say that
when the ``small scale" resistance of the elementary building block
increases, the tendency for localization at larger scales becomes stronger.

\section{Difficulties with the Landauer-type
formulation for the NSN conductance}

\label{LANDAUER}

The subtleties involved in producing a consistent Landauer-type
formula for the conductance of an NSN structure within the BDG
formulation are best explained by considering the simplest
single-mode, two-terminal configuration at zero temperature. In
other words,  we assume that there is no scattering in the system
except for the potential barriers at the interfaces, such that the
transverse channel modes are not mixed (and their indices can be
omitted).

We start with the purely normal case, as shown in the left panel of
Fig. \ref{SYS}, except that the S section is replaced by a normal
one (for example, by letting its gap approach zero). In this case
there is no need to treat electrons and holes concomitantly, as
there are no Andreev processes. It is enough to use only electrons
(or only holes). We denote the reflection coefficient for an
electron coming  from the left by ${\cal R}$, and the one for an electron
coming from the right by ${\cal R}'$. Likewise, the transmission
coefficient from the left to the right is denoted by ${\cal T}$, and the
one from the right to the left is ${\cal T}'$. Unitarity (particle
conservation, which is also charge conservation in this case)
implies
\begin{align}
{\cal R+T=R'+T'}=1\ .\label{UNITA}
\end{align}
Time-reversal symmetry implies further that ${\cal T=T'}$, and hence
${\cal R=R'}$. Next we assign to the left conductor  a chemical potential
$\mu_{L}$ and to the right one  a chemical potential $\mu_{R}$. In
the linear response regime, $\mu_{L}-\mu_{R}\rightarrow 0$. The
middle conductor, of a finite length, is kept floating, (i.e., it
is not connected to any reservoir), and will acquire a chemical
potential $\mu_{n}$. Clearly, the right-going currents to the left
of the middle segment, $I_{L}$,  and to its right, $I_{R}$, are
given by \cite{BOOK}
\begin{align}
I_{L}&=\frac{2e^{}}{h}\Bigl ((1-{\cal R})(\mu_{L}-\mu_{n})
-{\cal T}'(\mu_{R}-\mu_{n})\Bigr )\ ,\nonumber\\
I_{R}&=\frac{2e^{}}{h}\Bigl (-(1-{\cal R}')(\mu_{R}-\mu_{n})
+{\cal T}(\mu_{L}-\mu_{n})\Bigr )\ . \label{CURLR}
\end{align}
From the unitarity condition (\ref{UNITA}) and time-reversal
symmetry, it follows that $I_{L}=I_{R}\equiv
I=(2e^{}/h){\cal T}(\mu_{L}-\mu_{R})$, {\em independently} of the value
of $\mu_{n}$ (which, in fact, drops out of the two equations). The
well-known Landauer formula for the conductance,
\begin{align}
G=(2e^{2}/h){\cal T}\ ,\label{LANDAU}
\end{align}
is immediately obtained, and is independent of $\mu_{n}$, as it
should be. \cite{KANE}

When the middle section is a superconductor, further Andreev-type
processes become possible. An electron can be reflected/transmitted
as a hole, and {\it vice versa.} For an electron incident from the
left, the probabilities for the Andreev reflection and transmission
processes  are denoted ${\cal R}_{\rm A}$ and ${\cal T}_{\rm A}$, respectively.
The corresponding quantities for an electron coming from the right
are ${\cal R}'_{\rm A}$ and ${\cal T}'_{\rm A}$. The unitary condition
(\ref{UNITA}) is now replaced by
\begin{align}
{\cal R+T}+{\cal R}_{\rm A}+{\cal T}_{\rm A}={\cal R'+T'}
+{\cal R'}_{\rm A}+{\cal T}'_{\rm A}=1\
.\label{UNITAS}
\end{align}
However, the charge conservation condition  now reads \cite{COM1}
\begin{align}
{\cal R+T}-{\cal R}_{\rm A}-{\cal T}_{\rm A}={\cal R'+T'}-
{\cal R'}_{\rm A}-{\cal T'}_{\rm A}=1\
.\label{CHS}
\end{align}
The two conditions, Eqs. (\ref{UNITAS}) and (\ref{CHS}), are {\em not}
compatible whenever the Andreev probabilities do not vanish, {\em except} for
a ring geometry (where their consistency is enforced). Moreover, while Eq.
(\ref{UNITAS}) always holds for the solutions of the BDG equation, Eq.
(\ref{CHS}) does not!

The expressions for the currents [see Eqs. (\ref{CURLR})] now
become (note that group velocity of the holes is opposite to that
of the electrons)
\begin{widetext}
\begin{align}
I_{L}&=\frac{2e^{}}{h}\Bigl ((1-{\cal R}+{\cal R}_{\rm A})(\mu_{L}-\mu_{S})
-({\cal T'}-{\cal T}'_{\rm A})(\mu_{R}-\mu_{S})\Bigr )\ ,\nonumber\\
I_{R}&=\frac{2e^{}}{h}\Bigl (-(1-{\cal R}'+{\cal R}'_{\rm A})(\mu_{R}-\mu_{S})
+({\cal T}-{\cal T}_{\rm A})(\mu_{L}-\mu_{S})\Bigr )\ ,\label{EQJAP}
\end{align}
\end{widetext}
where $\mu_{S}$ is the chemical potential on the superconductor. We note that
because charge conservation does not hold [see Eq. (\ref{CHS})], $\mu_{S}$
does not drop out of these equations. Its value is relevant. Equations
(\ref{EQJAP}) are of the same form as Eqs. (2) of Takane and Ebisawa:
\cite{JAP} these authors determine $\mu_{S}$ so that $I_{L}=I_{R}$. It is
then possible to obtain a conductance from the current-to voltage ratio, as
was done in Ref. ~\onlinecite{JAP}. We reproduce their result for the NSN
conductance  in Sec. \ref{DISCA}, see Eq. (\ref{ISOF}) there. It is
disturbing, however, that the determined value of $\mu_{S}$ is relevant.
Moreover, for the result for the conductance [Eq. (\ref{ISOF})] to be fully
satisfactory, it should agree with the linear response, Kubo, formula for the
related  large ring geometry. This is the case for the normal Landauer
formula, but, in general not for Eq. (\ref{ISOF}) (see Sec. \ref{DISCA} for
more details).

The above considerations, in particular, Eqs. (\ref{UNITAS}) and
(\ref{CHS}), can be put on a more general basis. By imposing the
appropriate boundary conditions on the plane-wave solutions of the
BDG equation it is possible  to derive the scattering matrix,
${\cal S}$, of the NSN structure. This (4$\times$4) matrix relates
the amplitudes of the incoming waves to those of the outgoing
ones, (see left panel in Fig. \ref{SYS})
\begin{align}
c^{}_{\rm out}={\cal S}c^{}_{\rm in}\ .\label{SCAT}
\end{align}
Here, \cite{BEENAKKER} the incoming amplitudes are
\begin{align}
c^{}_{\rm in}=\Bigl (c^{+}_{e}(N),c^{-}_{e}(N'),c^{-}_{h}(N),c^{+}_{h}(N')\Bigr )\ ,
\label{IN}
\end{align}
and the outgoing ones are
\begin{align}
c^{}_{\rm out}=\Bigl (c^{-}_{e}(N),c^{+}_{e}(N'),c^{+}_{h}(N),c^{-}_{h}(N')\Bigr )\ .
\label{OUT}
\end{align}
In Eqs. (\ref{IN}) and (\ref{OUT}), $c^{+}_{e,h}(N)$ denotes the
amplitude of an electron-like (hole-like) excitation with a
positive wave vector $k_{e,h}
$ incident from the left normal side  while $c^{-}_{e,h}(N)$
refers to the waves having negative wave vectors. Since the BDG equation
conserves the number of quasiparticles, the scattering matrix
${\cal S}$ is necessarily unitary, and therefore
\begin{align}
c^{\dagger}_{\rm out }c^{}_{\rm out}=c^{\dagger}_{\rm in}c^{}_{\rm
in}\ .\label{NUM}
\end{align}
However,  conservation of the charge current \cite{BTK,COM1}
requires [see Eq. (\ref{CHS})]
\begin{align}
&|c^{+}_{e}(N^{}_{})|^{2}- |c^{-}_{e}(N^{}_{})|^{2}
+|c^{+}_{h}(N^{}_{})|^{2}-|c^{-}_{h}(N^{}_{})|^{2} \nonumber\\
&=|c^{+}_{e}(N'^{}_{})|^{2}- |c^{-}_{e}(N'^{}_{})|^{2}
+|c^{+}_{h}(N'^{}_{})|^{2}-|c^{-}_{h}(N'^{}_{})|^{2}\ .\label{CHA}
\end{align}
Comparing Eqs. (\ref{NUM}) and (\ref{CHA}), we see that they imply
\begin{align}
&|c^{+}_{e}(N_{}^{})|^{2}+|c^{-}_{e}(N'^{}_{})|^{2}=
|c^{+}_{e}(N')|^{2}+|c^{-}_{e}(N)|^{2}\ ,\nonumber\\
&|c^{-}_{h}(N)|^{2}+|c^{+}_{h}(N')|^{2}=
|c^{+}_{h}(N)|^{2}+|c^{-}_{h}(N')|^{2}\ .\label{BE}
\end{align}
Namely, the sum of the amplitudes squared of the incoming
electron-like excitations is equal to the sum of the amplitudes
squared of the outgoing electron-like excitations, and so is the
situation for the hole-like ones. These conditions are the same as those
 that would have been  derived from Eqs. (\ref{UNITAS}) and (\ref{CHS}), had we
 required that both conditions should be satisfied together. This always holds for
a normal system, in which these two types of quasiparticles are
not mixed. However, in a superconductor the Andreev processes mix
the hole-like with the electron-like excitations, thus violating
the conditions (\ref{BE})  for general hybrid structures with a
finite-size S segment. An exception is the ring geometry. There,
(see Fig. \ref{SYS}) the ratios $c^{+}_{e,h}(N)/c^{+}_{e,h}(N')$
are necessarily  phase factors, and so are the ratios
$c^{-}_{e,h}(N)/c^{-}_{e,h}(N')$. As a result, the plane-wave
solutions of the BDG equations for the ring geometry do satisfy
both conditions (\ref{BE}), namely, these solutions conserve
charge. One may therefore employ the BDG equation for the ring
geometry in the context of the Kubo formulation to
calculate the conductance of the NSN structure.

\section{The Kubo formula for a large ring}

\label{KUB}

For an infinite system, the Kubo-type conductivity at frequency
$\omega$ may be most easily obtained by calculating, using the
golden rule, the power absorbed by the system from a classical
monochromatic electromagnetic field. We consider for simplicity
noninteracting fermions (or Fermi quasiparticles), and focus on
the $\sigma_{xx}$ component of the conductivity,
\begin{align}
\sigma_{xx}^{}(\omega)&=-\frac{\pi e^{2}}{\rm
V}\frac{1}{\omega}\sum_{j,\ell}|\langle j|
v_{x}|\ell\rangle|^2\nonumber\\
&\times\delta(\epsilon_\ell-\epsilon_j-\hbar\omega)(f(\epsilon_j)-f(\epsilon_\ell))\
. \label{sigxx}
\end{align}
Here, $|j\rangle $ and  $|\ell\rangle$ are the quasi-electron
states and $f(\epsilon_{j})$ and $ f(\epsilon_\ell)$ their
populations. In Eq. (\ref{sigxx}), V is the volume of the system,
to be sent to infinity at the end of the calculation, at which
stage the summations over the states are replaced  by integrations
with the densities of states.  The $x-$component of the velocity
operator is denoted $ v_{x}$. For the case of a
normal (i.e., non superconducting) scatterer {\it with infinite
leads} the equivalence of the Kubo and the two-terminal Landauer
approaches has been established in Refs. ~\onlinecite{ES} and
~\onlinecite{FL}.

The assumption of an infinite system is crucial  in order to have
a continuum of states. An {\it isolated} finite system with a
truly discrete spectrum does {\it not} in fact  absorb
energy from the monochromatic field. In order to obtain a finite
conductivity for a finite large system, it has to be (and it is,
in most real situations) coupled to a very large heat bath. For
example, to an assembly of thermal phonons. This enables energy to
be transferred from the electromagnetic field into the bath via
the small electronic system. For a weak enough interaction with
the bath, one may say that the discrete levels of the system have
acquired finite widths, $\eta_j$. It then makes sense to
write down Eq. (\ref{sigxx}) with the levels having a finite width
(or with an imaginary part  to the frequency $\omega$, which will
amount to a non-monochromatic driving field). This procedure has
been discussed, including  the dc limit (${\rm
Re}\omega\rightarrow 0$), by Thouless and Kirkpatrick, \cite{TK}
following Czycholl and Kramer, \cite{CK} and used for example in
Ref. ~\onlinecite{IMRY}, see also Ref. ~\onlinecite{BOOK}. It is
postulated, and can be demonstrated in typical cases, that once
the $\eta_j$'s are larger than the level spacing near the Fermi
energy, but much smaller than all other relevant energy scales in
the problem, this procedure yields the physically relevant
low-frequency conductance of the system.

It hence follows that the $\omega\rightarrow 0$ conductance is obtained upon
transforming  the summations in Eq. (\ref{sigxx}) into energy integrations.
This allows one to approximate $f(\epsilon_{\ell}+\hbar\omega
)-f(\epsilon_{\ell})\simeq \hbar\omega f'(\epsilon_{\ell})$. Focusing on our
one-dimensional configuration, we take $x$ along the ring circumference, and
replace the volume of the system by its length, $d$. Since the
(one-dimensional) conductance is related to the conductivity by
$G\equiv\sigma /d$, we  recover the Kubo-Greenwood-type formula at low
frequencies
\begin{align}
G =\frac{e^{2}h}{2}  \nu ^{2} \sum_{\rm deg}
|\langle |v|\rangle |^{2} \ , \label{Kubos}
\end{align}
where the sum is over the (almost) degenerate initial states and over the
(almost) degenerate  final states,  within the narrow range
$\hbar\omega~(\rightarrow 0)$ above those initial states, and all states are
at about the Fermi energy. One might also say that the cancellation of the
frequency [see Eq. (\ref{sigxx})] is caused by the fact that the initial
state was, at $T=0$, within $\hbar \omega$ of the Fermi level. In Eq.
(\ref{Kubos}), the matrix element squared of the velocity was replaced by its
typical value in the small relevant energy window around the Fermi energy.
The double sum of Eq. (\ref{sigxx}) gave rise to two factors of the
single-particle density of states (per unit energy, per unit length, and per
spin), $\nu$,
\begin{align}
\nu =1/(hv_{\rm F})\ ,
\end{align}
in the one-dimensional system. Comparing Eq. (\ref{Kubos}) with
the ``traditional" Landauer formula (\ref{LANDAU}), we find that
in the Kubo approach the total transmission is replaced by the
appropriate sum over the velocity matrix elements squared, i.e.,
\begin{align}
G=\frac{2e^{2}}{h}\frac{1}{4v^{2}_{\rm F}}\sum_{\rm deg}
|\langle |v|\rangle |^{2}\ .\label{KUBA}
\end{align}

It is instructive to review the way  the Kubo formula in its form
(\ref{KUBA}) produces the Landauer result for the usual two-probe geometry.
We consider initial left-going scattering states. These are degenerate with
the right-going ones. This degeneracy gives a factor of two in the final
result, to which the spin degeneracy adds another factor of two. Each such
state will have a matrix element,   $(1-{\cal R+T})/2 = {\cal T}$, with the
appropriate final left-going scattering state, and $rt$ with the final
right-going scattering state ($r$ and $t$ are the reflection and transmission
amplitudes, respectively).  Adding the absolute values squared together
yields ${\cal T}$ (note the cancellation of the ${\cal T}^2$ term!).
Introducing the above degeneracy factors gives $\sum_{\rm deg} |\langle
|v|\rangle |^{2} =4 v_{F}^{2}{\cal T}$, and thus reproduces the Landauer
result, Eq. (\ref{LANDAU}) above.

For the ring geometry the states are
stationary and normalizable. Taking as a representative
example a normal ring with a single delta-function potential,
one finds that
the ratio of the  amplitudes of the clockwise moving wave
and anticlockwise moving one is a phase factor, $\exp [i\phi_{e}]$,
where on the Fermi energy
$\exp [i\phi_{e}]=-1$ or
$(1-i\zeta)/(1+i\zeta)$. Here $\zeta$ is the strength of the delta function potential,
with the corresponding transmission ${\cal T}=1/(1+\zeta^{2})$.
The velocity matrix elements are then $v_{\rm F}/(1\pm i\zeta )$, and thus
together with the spin degeneracy reproduce the Landauer formula (\ref{LANDAU}).
We give more details in the next section, which is devoted to the
evaluation of the states
and the current matrix elements for an NS ring, and the
case of an entirely normal ring is treated as a limiting case.

\section{The velocity matrix elements}

\label{MAIN}

Here we compute the matrix elements of the velocity operator,
which are used in the Kubo formula (\ref{KUBA}) for the
conductance. We follow BTK in assuming that the entire scattering
takes place only at the NS interfaces, and that the pair-potential
$\Delta$ is finite and spatially-invariant in the superconducting
region, and vanishes in the normal one. Since then there is no
channel mixing, the problem becomes effectively one-dimensional,
and the quasiparticles are  described by the one-dimensional
Bogoliubov-DeGennes equation, (we use in this section units in
which $\hbar =1$)
\begin{align}
\left [\begin{array}{cc}-\frac{1}{2m}\frac{d^{2}}{dx^{2}}-E_{\rm
F}&\Delta \\
\Delta &\frac{1}{2m}\frac{d^{2}}{dx^{2}}+E_{\rm F}
\end{array}\right ]\Psi (x)=\epsilon\Psi (x)\ .\label{BDG}
\end{align}
Note that this equation takes into account the two possible spin
directions pertaining to a certain energy $\epsilon$ (measured
from the Fermi level). In N, where $\Delta =0$, the solutions of
Eq. (\ref{BDG}) are
\begin{align}
\Psi^{\pm}_{e}(x)=\left [\begin{array}{c}1\\
0\end{array}\right ]e^{\pm ik_{e}x}\ ,\ \
\Psi^{\pm}_{h}(x)=\left [\begin{array}{c}0\\
1\end{array}\right ]e^{\pm ik_{h}x}\ ,\label{NNSOL}
\end{align}
with the wave vectors
\begin{align}
k_{e,h}^{}=\sqrt{2m (E_{\rm F}\pm \epsilon )}\simeq k_{\rm
F}\pm\epsilon/v_{\rm F}\ .\label{keh}
\end{align}
In the superconducting segment the solutions are
\begin{align}
\Psi^{\pm}_{e}(x)=\left [\begin{array}{c}\tilde{u}\\
\tilde{v}\end{array}\right ]e^{\pm iq_{e}x}\ ,\ \
\Psi^{\pm}_{h}(x)=\left [\begin{array}{c}\tilde{v}\\
\tilde{u}\end{array}\right ]e^{\pm iq_{h}x}\ .\label{STS}
\end{align}
Here,
\begin{align}
q_{e,h}^{}=\sqrt{2m (E_{\rm F}\pm\Omega )}\simeq k_{\rm
F}\pm\Omega/v_{\rm F}\ ,\label{qeh}
\end{align}
where
\begin{align}
\Omega &=\sqrt{\epsilon^{2}-\Delta^{2}}\ ,\ \ \epsilon \geq\Delta\
,\nonumber\\
\Omega &=i\sqrt{\Delta^{2}-\epsilon^{2}}\ ,\ \ \epsilon\leq\Delta\
,\label{OMEG}
\end{align}
and
\begin{align}
\tilde{u}^{2}=\frac{\epsilon}{\Omega}\Bigl
(\frac{1}{2}+\frac{\Omega}{2\epsilon }\Bigr )\ ,\ \ \
\tilde{v}^{2}=\frac{\epsilon}{\Omega}\Bigl
(\frac{1}{2}-\frac{\Omega}{2\epsilon }\Bigr )\ .\label{UV}
\end{align}
The factor $[\epsilon /\Omega]^{1/2}$ compensates  for the different
group velocity of the quasiparticles in the superconductor
[$\partial\epsilon /\partial q=(q/m)(\Omega /\epsilon )$], and it
multiplies  the usual coherence factors  $u$ and $v$,
$\tilde{u}=[\epsilon /\Omega]^{1/2}u$ and $\tilde{v}=[\epsilon
/\Omega]^{1/2}v$. The amplitudes $c^{\pm}_{e,h}(N)$ (see Fig.
\ref{SYS}) are the coefficients of the waves $\exp [\pm ik_{e,h}x]$.
Analogous amplitudes are defined for the waves in  the
superconducting segment.

For simplicity, we assume that the  (left) NS interface at $x=0$
(see Fig. \ref{SYS})  is  represented by a delta-function
potential, $\lambda \delta (x)$, of strength $\zeta =\lambda
/v_{\rm F}$. Then the boundary conditions \cite{COM1} are the
continuity of the wave functions and the discontinuity (of
magnitude $\zeta $) of their derivatives, leading to the following
relations among the amplitudes of the N region and those of the S
one,
\begin{widetext}
\begin{align}
&\left [\begin{array}{c}c^{+}_{e}(S)\\ c^{-}_{e}(S)\\
c^{+}_{h}(S)\\ c^{-}_{h}(S)\end{array}\right ]= \left
[\begin{array}{cccc}\tilde{u}&0&-\tilde{v}&0\\
0&\tilde{u}&0&-\tilde{v}\\
-\tilde{v}&0&\tilde{u}&0\\
0&-\tilde{v}&0&\tilde{u}\end{array}\right ]\left [\begin{array}{cccc}1-i\zeta &-i\zeta &0&0\\
i\zeta &1+i\zeta &0&0\\
0&0&1-i\zeta &-i\zeta \\
0&0&i\zeta &1+i\zeta \end{array}\right ]\left [\begin{array}{c}c^{+}_{e}(N)\\ c^{-}_{e}(N)\\
c^{+}_{h}(N)\\ c^{-}_{h}(N)\end{array}\right ] \ .\label{BC1}
\end{align}
The other NS interface, located at $x=d_{S}$, is assumed to be
perfectly transparent, and then the boundary conditions are the
continuity of the wave functions and their derivatives. When the
system has the shape of a ring, in which the length of the normal
segment is $d_{N}$, these boundary conditions are
\begin{align}
\left
[\begin{array}{c}e^{-ik_{\rm F}d_{N}}\gamma_{N}^{-1}c^{+}_{e}(N)\\
e^{ik_{\rm F}d_{N}}\gamma_{N}^{}c^{-}_{e}(N)\\
e^{-ik_{\rm F}d_{N}}\gamma_{N}^{}c^{+}_{h}(N)
\\
e^{ik_{\rm F}d_{N}}\gamma_{N}^{-1}c^{-}_{h}(N)\end{array}\right
]=\left [\begin{array}{cccc}e^{ik_{\rm
F}d_{S}}\tilde{u}\gamma_{S}^{}
&0&e^{ik_{\rm F}d_{S}}\gamma_{S}^{-1}\tilde{v}&0\\
0&e^{-ik_{\rm F}d_{S}}\tilde{u}\gamma_{S}^{-1}
&0&e^{-ik_{\rm F}d_{S}}\gamma_{S}^{}\tilde{v}\\
e^{ik_{\rm F}d_{S}}\tilde{v}\gamma_{S}^{}
&0&e^{ik_{\rm F}d_{S}}\gamma_{S}^{-1}\tilde{u}&0\\
0&e^{-ik_{\rm F}d_{S}}\tilde{v}\gamma_{S}^{-1} &0&e^{-ik_{\rm
F}d_{S}}\gamma_{S}^{}\tilde{u}\end{array}\right ]
\left [\begin{array}{c}c^{+}_{e}(S)\\ c^{-}_{e}(S)\\
c^{+}_{h}(S)\\ c^{-}_{h}(S)\end{array}\right ]\ ,\label{BC2}
\end{align}
\end{widetext}
where
\begin{align}
\gamma_{N}=e^{i\epsilon d_{N}/v_{\rm F}}\ ,\ \ \gamma_{S}=e^{i\Omega
d_{S}/v_{\rm F}}\ .\label{gama}
\end{align}
Without loss of generality we may choose $\exp (ik_{\rm F}d)=1$,
where $d=d_{N}+d_{S}$ is the total length of the ring. Then $k_{\rm
F}$ disappears from the boundary conditions.

Upon eliminating the S-region amplitudes, one obtains the equation
which determines the allowed eigenenergies of the ring, and the
ratios among the amplitudes of the normal region for each such
energy,
\begin{align}
&\Biggl (\left
[\begin{array}{cccc}X&0&Y&0\\
0&X^{\ast}&0&Y^{\ast}\\
Y^{\ast}&0&X^{\ast}&0\\
0&Y&0&X\end{array}\right
]\nonumber\\
&-\left [\begin{array}{cccc}1-i\zeta &-i\zeta &0&0\\
i\zeta &1+i\zeta &0&0\\
0&0&1-i\zeta &-i\zeta \\
0&0&i\zeta &1+i\zeta \end{array}\right ]\Biggl )
\left [\begin{array}{c}c^{+}_{e}(N)\\ c^{-}_{e}(N)\\
c^{+}_{h}(N)\\ c^{-}_{h}(N)\end{array}\right ]=0\ .\label{HAMATA}
\end{align}
Here,
\begin{align}
X&=\gamma_{N}^{-1}\Bigl (\cos (\Omega d_{S}/v_{\rm
F})-i\frac{\epsilon}{\Omega}\sin
(\Omega d_{S}/v_{\rm F})\Bigr )\ ,\nonumber\\
Y&=2i \gamma_{N}^{}\tilde{u}\tilde{v}\sin (\Omega d_{S}/v_{\rm F})\
, \label{XY}
\end{align}
such that $|X|^{2}-|Y|^{2}=1$ for both  $\epsilon\geq\Delta$
and $\epsilon\leq\Delta$. The allowed eigenenergies are given by the
vanishing of the determinant of the matrix in Eq. (\ref{HAMATA}).
The  zeroes of the determinant define the families of possible
eigenenergies, which are rather dense when the size of the entire
system is large. When $\zeta\neq 0$, the eigenvectors of the matrix
(\ref{HAMATA}) are such that the ratios of the clockwise electron
(hole) waves to the anticlockwise electron (hole) ones (see Fig.
\ref{SYS}) for each of the families of eigenenergies are phase
factors,
\begin{align}
\frac{c^{-}_{e}(N)}{c^{+}_{e}(N)}=e^{i\phi_{e}}\ ,\ \
\frac{c^{-}_{h}(N)}{c^{+}_{h}(N)}=e^{i\phi_{h}}\ .\label{PF}
\end{align}
In other words, at any finite value $\zeta$ of the barrier, there is
a perfect reflection of the electron and the
hole waves (in the ring geometry). On the
other hand, the ratios of the hole amplitudes to the electron ones
obey
\begin{align}
\frac{c^{+}_{h}(N)}{c^{+}_{e}(N)}&=P_{}^{}\ ,\ \
\frac{c^{-}_{h}(N)}{c^{-}_{e}(N)}=P_{}^{\ast}\ ,\label{P}
\end{align}
such that the phase of $P_{}$ is $(\phi_{e}-\phi_{h})/2$,
\begin{align}
P_{}^{}=|P_{}^{}|e^{i(\phi_{e}-\phi_{h})/2}\ .\label{PPHASE}
\end{align}

It is illuminating to consider   Eq. (\ref{HAMATA}) and its
solutions in the limit of very high energies, $\epsilon\gg\Delta$,
where the superconducting order parameter $\Delta$ becomes
irrelevant, and the entire system behaves as if it were normal.
Then [see Eqs. (\ref{XY})] $X=\exp [i\epsilon d/v_{\rm F}]$ and
$Y=0$, and Eq. (\ref{HAMATA}) separates into two independent
blocks, for the electron-like excitations, and for the hole-like
ones. The eigenenergies are given by
\begin{align}
&e^{i\epsilon d/v_{\rm F}}=1 \ {\rm or}\ \frac{1+i\zeta}{1-i\zeta}\
,\ {\rm for}\ {\rm the}\
{\rm electron}\ {\rm waves}\ ,\nonumber\\
&e^{i\epsilon d/v_{\rm F}}=1\ {\rm or}\ \frac{1-i\zeta}{1+i\zeta}\
,\ {\rm for}\ {\rm the}\ {\rm hole}\ {\rm waves}\ ,\label{HIE}
\end{align}
with the corresponding phase ratios
\begin{align}
e^{i\phi_{e}}&=e^{i\phi_{h}}=-1\ ,\ {\rm or}\ \
\frac{1-i\zeta}{1+i\zeta}\ .\label{RATIOH}
\end{align}
(In this limit $P$, the ratio of the hole amplitude to the
electron amplitude, is not defined.) We show below that these are
the phase factors $\exp [i\phi_{e,h}]$ which determine the
conductance (\ref{LANDAU}) when calculated from the Kubo formula
(\ref{KUBA}).

In the other extreme limit of sub-gap energies, $\epsilon\ll\Delta$,
one approximates \cite{BTK} [see Eq. (\ref{OMEG})]
\begin{align}
\frac{\Omega}{v_{\rm F}}\simeq \frac{\Delta}{v_{\rm
F}}\equiv\frac{1}{\xi}\ ,
\end{align}
and consequently [see Eqs. (\ref{XY})]
\begin{align}
X\simeq \gamma_{N}^{-1}{\rm cosh}(d_{S}/\xi )\ ,\ \ Y\simeq
-i\gamma_{N}^{}{\rm sinh} (d_{S}/\xi )\ ,
\end{align}
where $\xi$ is the coherence length in the superconductor. One then
finds a quadratic equation for $\cos (\epsilon d_{N}/v_{\rm F})$. We
do not present explicit expressions for the solutions and the
amplitude ratios since they are rather cumbersome.

The next step in this calculation is to find the normalization of
the wave functions, using
\begin{align}
\int_{0}^{d_{S}}dx|\Psi_{S}(x)|^{2}&+\int_{d_{S}}^{d_{}}dx|\Psi_{N}(x)|^{2}=
1\ .\label{FULLNOR}
\end{align}
In the N region
\begin{align}
&|\Psi_{N}(x)|^{2}=|c^{+}_{e}(N)|^{2}+|c^{-}_{e}(N)|^{2}\nonumber\\
&+(e^{2ik_{F}x}c^{-}_{e}(N))^{\ast}c^{+}_{e}(N)+{\rm cc})+({\rm
e}\rightarrow{\rm h})\ .
\end{align}
When $d_{N}$ is large, such that the oscillatory terms (in $k_{\rm
F}d_{N}$) can be ignored, the contribution of the normal part to the
normalization integral becomes
\begin{align}
\int_{d_{S}}^{d}dx|\Psi_{N}(x)|^{2}&=d_{N}\Bigl
(|c^{+}_{e}(N)|^{2}+|c^{-}_{e}(N)|^{2}\nonumber\\
&+|c^{+}_{h}(N)|^{2} +|c^{-}_{h}(N)|^{2}\Bigr ) \ .\label{norN}
\end{align}
The calculation of the contribution to the normalization coming from
the S region is more subtle, since the wave vectors can have an
imaginary part [see Eqs. (\ref{qeh}) and (\ref{OMEG})]. Disregarding
terms oscillating with  $k_{\rm F}d_{S}$, we find
\begin{align}
|\Psi_{S}(x)|^{2}&\rightarrow (|\tilde{u}|^{2}+|\tilde{v}|^{2})\Bigl
( (|c^{+}_{e}(S)|^{2}+|c_{h}^{-}(S)|^{2})e^{i\frac{(\Omega
-\Omega^{\ast})x}{v_{\rm F}}}\nonumber\\
& + (|c^{-}_{e}(S)|^{2}+|c_{h}^{+}(S)|^{2})e^{i\frac{(\Omega^{\ast}
-\Omega)x}{v_{\rm F}}}\Bigr )\nonumber\\
&+(\tilde{u}^{\ast}\tilde{v}+\tilde{u}\tilde{v}^{\ast})\Bigl
(e^{i\frac{(\Omega
+\Omega^{\ast})x}{v_{\rm F}}}((c^{-}_{e}(S))^{\ast}c^{-}_{h}(S)\nonumber\\
&+(c^{+}_{h}(S))^{\ast}c^{+}_{e}(S))+{\rm cc}\Bigr )\ .\label{PSI2}
\end{align}
This rather complicated  result reflects  the fact (specifically,
its second part) that in the superconductor the electron waves are
mixed with the hole ones. However, at very large energies,
$\epsilon\gg\Delta$, or at very small ones, $\epsilon\ll\Delta$,
the mixing term,
$(\tilde{u}^{\ast}\tilde{v}+\tilde{u}\tilde{v}^{\ast})$, vanishes.
In the following, we confine ourselves to these two limits. In the
high-energies limit  the normalization of either the clockwise
waves or the anticlockwise ones is simply $\sqrt{2d}$, where $d$
is the total length. In the low-energies limit we  find
\begin{align}
\int_{0}^{d_{S}}dx|\Psi_{S}(x)|^{2}&=\frac{\xi}{2}{\rm
sinh}(d_{S}/\xi)
(|c^{+}_{e}(N)|^{2}+|c^{-}_{e}(N)|^{2})\nonumber\\
&\times\Bigl (e^{d_{S}/\xi }|M^{}_{a}|^{2}
+e^{-d_{S}/\xi}|M^{}_{b}|^{2} \Bigr )\ ,
\end{align}
where
\begin{align}
M^{}_{a}&=\gamma_{N}^{-1}-i\gamma_{N}^{}P\ ,\nonumber\\
M^{}_{b}&=\gamma_{N}^{-1}+i\gamma_{N}^{}P\ ,\label{MAMB}
\end{align}
and $P$ is given by Eq. (\ref{P}).

Having fully determined the wave functions, it remains to compute
the matrix elements of the velocity,
\begin{align}
v^{}_{j\ell}\equiv\langle j|v| \ell\rangle=\frac{1}{2mi}
\int_{0}^{d} dx \Bigl (\Psi^{\ast}_{j}\frac{d\Psi^{}_{\ell}}{dx}-
\Psi^{}_{\ell}\frac{d\Psi^{\ast}_{j}}{dx}\Bigr ) \ ,\label{IJL}
\end{align}
with the indices $j$ and $\ell$ enumerating the various
eigenfunctions. As in the calculation of the normalization, here
again there are contributions from the normal and from the
superconducting regions. In each region we discard the oscillatory
terms, those which involve $k_{\rm F}d_{N}$ or $k_{\rm F}d_{S}$.

The contribution of the normal part to the integral in Eq.
(\ref{IJL}) reads
\begin{align}
v^{N}_{j\ell}&=\frac{d_{N}k_{\rm F}}{m}\Bigl
((c^{+}_{e}(N))^{\ast}_{j} (c^{+}_{e}(N))^{}_{\ell} -
(c^{-}_{e}(N))^{\ast}_{j}(c^{-}_{e}(N))^{}_{\ell}\nonumber\\
& + (c^{+}_{h}(N))_{j}^{\ast}(c^{+}_{h}(N))^{}_{\ell}-
(c^{-}_{h}(N))_{j}^{\ast}(c^{-}_{h}(N))^{}_{\ell} \Bigr )\ .
\label{IINN}
\end{align}
In the high-energies limit, $\epsilon\gg\Delta$, the contribution
of the superconducting segment to the integration is the same as
(\ref{IINN}) (with the arguments $N$ replaced by  $S$,  and
$d_{N}$ replaced by $d_{S}$). The contribution of the S part  in
the limit of very low energies, $\epsilon\ll\Delta$, is
\begin{align}
&v^{S}_{j\ell}=\frac{\xi k_{\rm F}}{2m}{\rm sinh}(d_{S}/\xi
)\nonumber\\
&\times\Bigl (
(c^{+}_{e}(N))^{\ast}_{j}(c^{+}_{e}(N))^{}_{\ell}M^{}_{j\ell}
-(c^{-}_{e}(N))^{\ast}_{j}(c^{-}_{e}(N))^{}_{\ell}M^{\ast}_{j\ell}\Bigl
)\ ,
\end{align}
where we have denoted [see Eqs. (\ref{MAMB})]
\begin{align}
M_{j\ell}^{}=e^{d_{S}/\xi}
(M^{}_{a})^{\ast}_{j}(M^{}_{a})^{}_{\ell}+
e^{-d_{S}/\xi}(M^{}_{b})^{\ast}_{j}(M^{}_{b})^{}_{\ell}\ .
\end{align}

It is again useful to examine the limit of high energies,
$\epsilon\gg\Delta$, where the entire ring behaves as if it were
normal. Then, the electron- and the hole-like waves are separated.
The spectrum and the amplitude ratios are given by Eqs.
(\ref{HIE}) and (\ref{RATIOH}), and the normalization for each
species is $\sqrt{2d}$. The matrix elements of the velocity are
simply
\begin{align}
v^{\rm ele}_{j\ell }=\frac{k_{\rm F}d}{m}(c^{+}_{e})^{\ast}_{j}
(c^{+}_{e})^{}_{\ell}(1- e^{i(\phi^{\ell}_{e}-\phi^{j}_{e})})\ ,\label{NORMALVEL}
\end{align}
and an analogous result is obtained for the contribution of the hole waves.
Obviously, the diagonal ones vanish. The non-diagonal ones give $(k_{\rm
F}/m)/(1\pm i\zeta )$,  and consequently
\begin{align}
\sum_{\rm deg}|\langle |v^{}|\rangle |^{2}=4v^{2}_{\rm F}{\cal T}\
, \ \ {\cal T}=\frac{1}{1+\zeta^{2}}\ .\label{INORMAL}
\end{align}
Note that the non-vanishing matrix elements arise from the phase
factor between waves belonging to the same species but moving along
opposite directions. Hence, the Kubo formulation for the ring
geometry reproduces the Landauer result for the dc conductance.

Another illuminating limit is when $\zeta$ vanishes, and both NS
interfaces (see Fig. \ref{SYS}) are perfectly transparent, in
which case the clockwise and the anticlockwise amplitudes are
independent. The  matrix elements of the velocity, for sub-gap
energies, are (for either the clockwise-moving  or the
anticlockwise-moving excitations)
\begin{align}
v^{}_{\ell\ell}&=v_{\rm F}\ ,\nonumber\\
v^{}_{j\ell}&=v^{\ast}_{\ell j}=v_{\rm F}\frac{d_{N}}{d_{N}+\xi
{\rm
tanh}(d_{S}/\xi )}\nonumber\\
&\times\frac{{\rm sinh}(d_{S}/\xi )[{\rm sinh}(d_{S}/\xi )+i]}{{\rm
cosh}^{2}(d_{S}/\xi )}\ .
\end{align}
(It is interesting to note that the off-diagonal matrix elements
are coming from the N region alone.) Hence the contribution of
both the clockwise waves and the anticlockwise waves is
\begin{align}
\sum_{\rm deg}|\langle |v^{}|\rangle |^{2}=4v^{2}_{\rm F}\Bigl
(1+\Bigr [\frac{d_{N}{\rm tanh} (d_{S}/\xi )}{d_{N}+\xi{\rm
tanh}(d_{S}/\xi )}\Bigr ]^{2}\Bigl )\ . \label{ITT}
\end{align}
Thus, when $d_{S}/\xi$ tends to zero (namely, in the absence of
the superconductor) the result approaches the Landauer formula for
a transparent barrier, {\it cf.} Eq. (\ref{INORMAL}). On the other
hand, when $d_{N}\geq d_{S}\gg\xi$, our result (\ref{ITT}) tends
to the one found by BTK  \cite{BTK} (for a clean interface),
namely, it is {\em twice} the value of the quantum conductance.

\begin{widetext}

\begin{figure}[h]
\includegraphics[width=8.cm,height=5.cm]{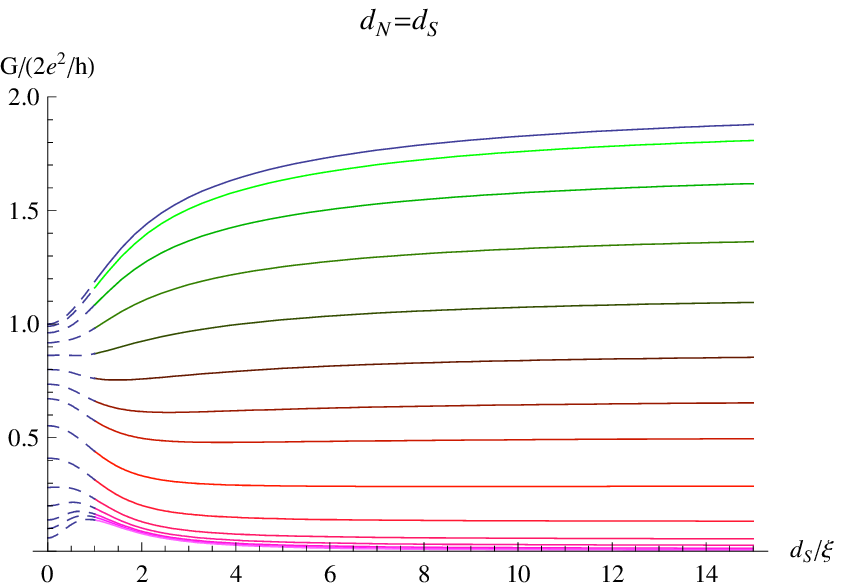}
\includegraphics[width=8.cm,height=5.cm]{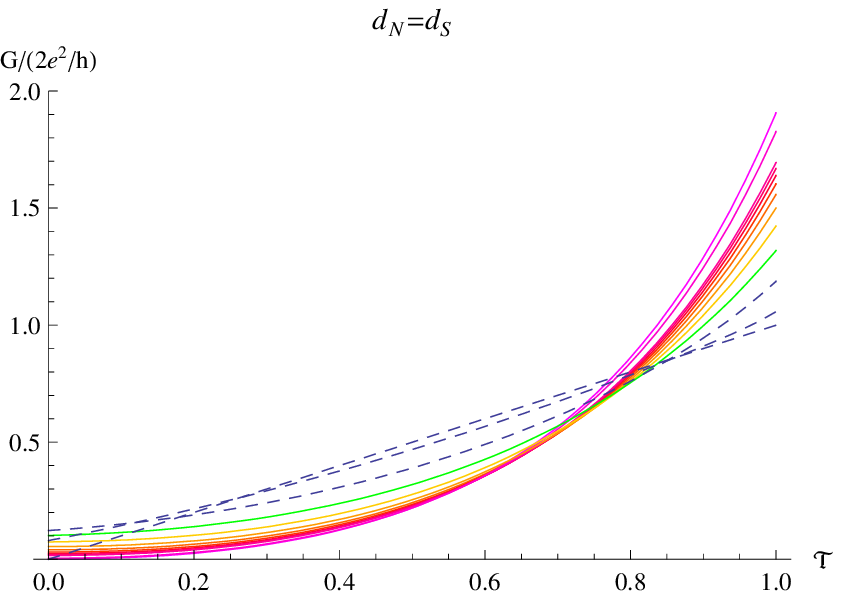}
\caption{Left panel: Conductance {\it vs} $d_{S}/\xi$ for several values of
the transmission. Right panel: Conductance {\it vs} the transmission for
several values of $d_{S}/\xi$. Here the length of the S region equals that of
the N region  ($d_{N}=d_{S}$).  All information pertaining to values of
$d_{S}$ smaller than $\xi$ is presented by dashed curves.} \label{EQ}
\end{figure}

\end{widetext}

\begin{widetext}

\begin{figure}[h]
\includegraphics[width=8cm,height=5.cm]{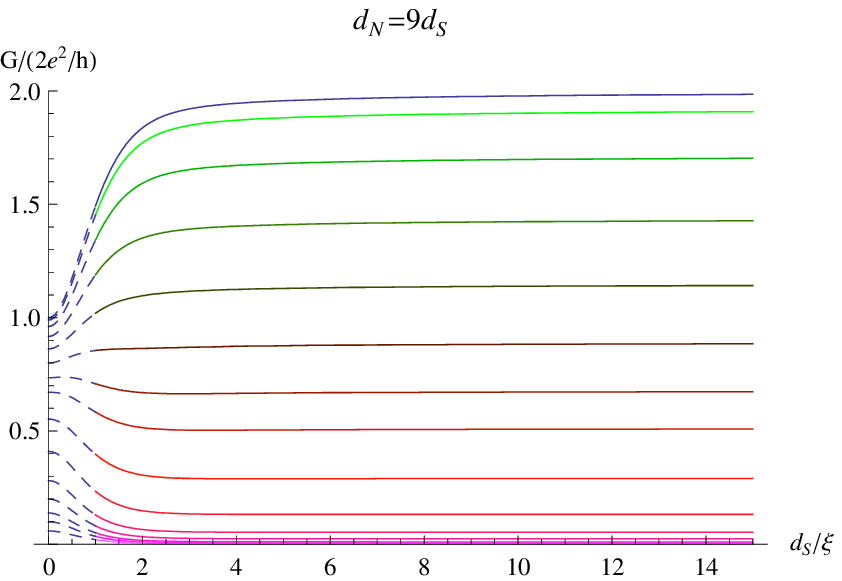}
\includegraphics[width=8cm,height=5.cm]{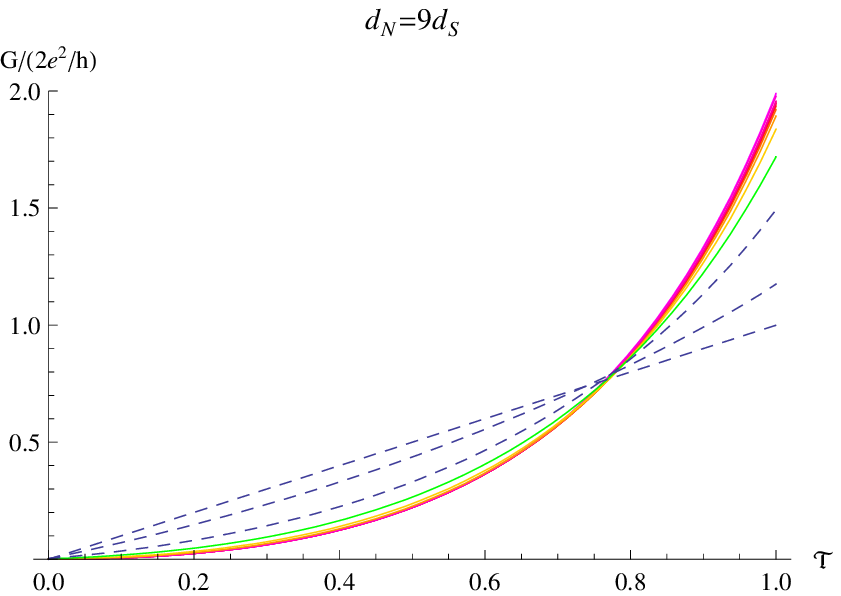}
\caption{Left panel: Conductance {\it vs} $d_{S}/\xi$ for several
values of the transmission. Right panel: Conductance {\it vs} the
transmission for several values of $d_{S}/\xi$.  Here the length of
the S region is smaller than the length of the N region
($d_{N}/d_{S}=9$). All information pertaining to values of $d_{S}$
smaller than $\xi$ is presented by dashed curves.} \label{NEQ}
\end{figure}

\end{widetext}

Unfortunately, the explicit expressions for the velocity matrix
elements at low energies for general values of $\zeta$ and
$d_{S}/\xi$ are rather complicated. Consequently, we present the
results   of the calculations only graphically, see  Figs.
\ref{EQ} and \ref{NEQ}. The figures show the conductance [divided
by 2$e^{2}/h$, see Eq. (\ref{KUBA})] as a function of the ratio
$d_{S}/\xi$ for various values of the interface transmission,
${\cal T}=1/(1+\zeta^{2})$, and as a function of that
transmission, for various values of the size of the superconductor
segment, $d_{S}/\xi$.

Figure \ref{EQ} presents the results for the case $d_{N}=d_{S}$,
i.e., the segments N and S of the ring are of equal lengths. In
the left panel the conductance is plotted as a function of
$d_{S}/\xi$, for values of ${\cal T}$ ranging between 1 (the
upmost curve) and 0.06, (the lowest-lying one). The main feature
of these curves is the variation of their slope as the NS barrier
becomes less and less transparent. For ${\cal T}=1$ the
conductance increases with the length of the superconductor (until
it is double that of the normal system in the BTK limit where
$d_{S}/\xi\rightarrow\infty$). As the transparency decreases, the
conductance, albeit increasing with $d_{S}/\xi$ becomes smaller,
until at about ${\cal T}\simeq 0.8$ it changes its slope and
begins {\em deceasing} as the size of the superconductor is
increased. The same characteristic behavior is obtained when the
size of the normal part largely exceeds that of the
superconductor, as is depicted in the left panel of Fig.
\ref{NEQ}. The right panels in both Figs. \ref{EQ} and \ref{NEQ} show the
(normalized) conductance as a function of the barrier transparency
for various values of the superconducting size, $d_{S}/\xi$,
ranging between 0.01 (almost a straight line) and 20 (parabolic curve). Here one
observes that the conductance is linear in the barrier
transmission as long as the superconducting is small enough, and
then becomes quadratic in ${\cal T}$, for large values of
$d_{S}/\xi$. It should be noted, however, that the use of the BDG
approach for $d_{S}/\xi\ll 1$ is dubious. For this region, we have
presented all information pertaining to such values by the dotted
curves.

\section{Discussion}

\label{DISCA}

As is described in Secs. \ref{INTROD} and \ref{LANDAUER},
several previous calculations
aiming to determine
the conductance of hybrid normal-superconducting structures are
based on the scattering matrix for the quasiparticles, as derived
from the BDG equation. \cite{JAP,LAMBERT,ANANTRAM} Our
reservations regarding this procedure are explained in Sec.
\ref{LANDAUER}. Nonetheless, it is interesting to compare the
conductance found from the Kubo formula and the one derived after
fixing the chemical potential of the superconductor, as explained
in Sec. \ref{LANDAUER}. Here we carry out this comparison for the
model system of Sec. \ref{MAIN}.

The scattering matrix  of the NSN junction (see the left panel in
Fig. \ref{SYS}) is a function of the energy $\epsilon$.
For our purposes here it suffices to
derive it for zero energy, i.e. on the Fermi level. This derivation
is accomplished by eliminating the amplitudes of the waves within
the superconductor, using the boundary conditions (\ref{BC1}), and
the boundary conditions at the (clean) interface between the
superconductor and the second normal layer, denoted N' [note that
when $\epsilon =0$, $\gamma_{N}=1$, see Eq. (\ref{gama})]
\begin{align}
\left
[\begin{array}{c}c^{+}_{e}(N')\\
c^{-}_{e}(N')\\
c^{+}_{h}(N')
\\
c^{-}_{h}(N')\end{array}\right ]=\left
[\begin{array}{cccc}\tilde{u}\gamma_{S}^{}
&0&\gamma_{S}^{-1}\tilde{v}&0\\
0&\tilde{u}\gamma_{S}^{-1}
&0&\gamma_{S}^{}\tilde{v}\\
\tilde{v}\gamma_{S}^{}
&0&\gamma_{S}^{-1}\tilde{u}&0\\
0&\tilde{v}\gamma_{S}^{-1}
&0&\gamma_{S}^{}\tilde{u}\end{array}\right ]
\left [\begin{array}{c}c^{+}_{e}(S)\\ c^{-}_{e}(S)\\
c^{+}_{h}(S)\\ c^{-}_{h}(S)\end{array}\right ]\ .\label{BC22}
\end{align}
As a result, the scattering matrix as defined in Eq. (\ref{SCAT})
takes the form
\begin{widetext}
\begin{align}
\left
[\begin{array}{c}c^{-}_{e}(N)\\
c^{+}_{e}(N')\\
c^{+}_{h}(N)
\\
c^{-}_{h}(N')\end{array}\right ]=\frac{1}{D}\left [\begin{array}
{cccc}-i\zeta (1-i\zeta )(c^{2}+s^{2})&(1-i\zeta )c&isc&-\zeta s\\
c(1-i\zeta )&-i\zeta (1-i\zeta )&\zeta s&isc (1+2\zeta^{2})\\
isc&\zeta s&i\zeta (1+i\zeta )(c^{2}+s^{2})&c(1+i\zeta )\\
-\zeta s&isc (1+2\zeta^{2})&c(1+i\zeta )&i\zeta (1+i\zeta
)\end{array} \right ]\left
[\begin{array}{c}c^{+}_{e}(N)\\
c^{-}_{e}(N')\\
c^{-}_{h}(N)
\\
c^{+}_{h}(N')\end{array}\right ]\ ,\label{SS}
\end{align}
\end{widetext}
where
\begin{align}
D=(1+\zeta^{2})c^{2}+\zeta^{2}s^{2}\ ,
\end{align}
and  in order to shorten the notations we have denoted
\begin{align}
s\equiv {\rm sinh} (d_{S}/\xi )\ ,\ \ c\equiv {\rm cosh}(d_{S}/\xi
)\ .
\end{align}

Referring to the notations introduced in Sec. \ref{LANDAUER}, we find from Eq.
(\ref{SS}) that for an electron-like incident from the left
\begin{align}
{\cal R}&=\frac{\zeta^{2}(1+\zeta^{2})(c^{2}+s^{2})^{2}}{D^{2}}\ ,\ \ {\cal T}=\frac{c^{2}(1+\zeta^{2})}{D^{2}}\ ,\nonumber\\
{\cal R}_{A}&=\frac{s^{2}c^{2}}{D^{2}}\ ,\ \ \ \ \ \
{\cal T}_{A}=\frac{\zeta^{2}s^{2}}{D^{2}}\ ,\label{LOP}
\end{align}
while for an electron-like wave coming from the right
the corresponding probabilities are
\begin{align}
{\cal R}'&=\frac{\zeta^{2}(1+\zeta^{2})}{D^{2}}\ ,\ \ \
{\cal T}'=\frac{c^{2}(1+\zeta^{2})}{D^{2}}\ ,\nonumber\\
{\cal R}'_{A}&=\frac{s^{2}c^{2}(1+2\zeta^{2})^{2}}{D^{2}}\ ,\ \ \ {\cal T}'_{A}=\frac{\zeta^{2}s^{2}}{D^{2}}\ .\label{IMP}
\end{align}
It is easy to verify that the conditions for
quasiparticle-number conservation, Eq. (\ref{UNITAS}),
are obeyed by the probabilities
(\ref{LOP}) and (\ref{IMP}), since the scattering matrix is unitary;
Eqs. (\ref{CHS}) for the charge conservation are not obeyed.
Following Refs.~ \onlinecite{JAP} and ~\onlinecite{ANANTRAM},
current conservation is now imposed on
Eqs. (\ref{EQJAP}), leading to the determination
of the chemical potential on the superconductor. This leads to a linear relation
between $I_{L}$ and the chemical potential difference $\mu_{L}-\mu_{R}$,
which is identified as the conductance. Denoting the latter by $G^{}_{\rm sc}$, one has
\begin{align}
G^{}_{\rm
sc}=\frac{g^{}_{LL}g^{}_{RR}-g^{}_{LR}g^{}_{RL}}{g^{}_{LL}+g^{}_{RR}+
g^{}_{LR}+g^{}_{RL}}\ ,\label{ISOF}
\end{align}
where \cite{JAP}
\begin{align}
g^{}_{ij}=\frac{2e^{2}}{h}\Bigl (\delta^{}_{ij}-|{\cal
S}^{ee}_{ij}|^{2}+|{\cal S}^{he}_{ij}|^{2}\Bigr )\ .
\end{align}
Here, $i $ and $j$ refer to the two sides of the junction, say left
and right, and the superscripts $ee$ or $he$ refer to the particular
process. Thus for example, the 11  element of the matrix in Eq.
(\ref{SS}) is ${\cal S}_{LL}^{ee}$, while the 41 element is ${\cal
S}^{he}_{RL}$.

\begin{figure}[h]
\includegraphics[width=8cm]{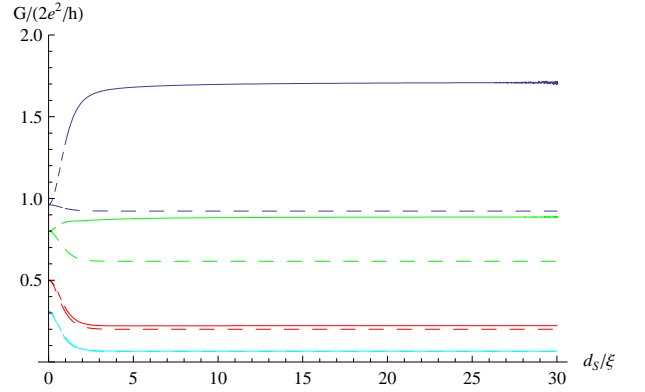}
\caption{Comparison between the conductance Eq. (\ref{ISOF}) (dashed curves)
and the conductance computed according to the Kubo formula, Eq. (\ref{KUBA}) (solid lines),
as a function of $d_{S}/\xi$ for four values of the interface transmission.
The latter conductance  is always larger than the former.} \label{HASS}
\end{figure}

We compare the outcome of Eq. (\ref{ISOF}) with the conductance found from
the Kubo formula in Fig. \ref{HASS}. There, the conductances are plotted for
four values of the interface transmission, an almost perfect one, ${\cal
T}=0.96$, (the upmost pair of curves), ${\cal T}=0.8$, (the second pair of
curves from above),  ${\cal T}=0.5$ and ${\cal T}=0.31$,  (the
low-lying two pairs of curves). In each case, the
result of Eq. (\ref{ISOF}) is the dashed line.  There are three interesting features
of this comparison. Firstly, the conductance found from Eq. (\ref{ISOF}) is
always {\em smaller} than the one found from the Kubo formula, Eq.
(\ref{KUBA}).  The difference between the two results decreases with
increasing barrier (decreasing $\cal T$) and seems to vanish in the limit
${\cal T} \rightarrow 0$. Thus, while for a normal conductor the ring
geometry and the simple two-terminal configuration produce identical results
for the conductance, this is unfortunately no longer the case for the NSN
junction (NS for the ring geometry). The second interesting feature concerns
the slopes of the curves in Fig. \ref{HASS}, when the barrier transmission,
${\cal T}$ is varied. While the conductance computed from the Kubo formula
shows a {\em crossover of the slope, from being positive at high values of
${\cal T}$ to being negative at lower values,} the slope of the conductance
found from Eq. (\ref{ISOF}) seems to be {\em always negative}. A third
important difference between the two approaches is that for large $d_{s}/\xi$
and not-too-small $\cal T$, the Kubo result becomes larger than $2e^{2}/h$
(tending to $4e^2 /h$ in the limit $d_{s}/\xi \rightarrow \infty$ {\em and}
${\cal T} \rightarrow 1$), while Eq. (\ref{ISOF}) actually tends to $G_{\rm NS} = 2e^2 /h$
and {\em never} yields the doubling of $G_{\rm NS}$ due to the Andreev reflections.
We blame these differences between the two approaches on
the lack of conservation of charge in the BDG formulation. We believe that
this deficiency is corrected by employing the Kubo formula for the ring
geometry.

The difference between the two
approaches becomes most marked in the limit of a nearly transparent barrier and a thick superconductor.
This can be easily understood
by noting that the addition of the two NS resistances is handled very
differently by the two approaches.
For the fully quantum case, adding two
ideal conductances ($4e^2/h$ for the NS case) gives just  one  ideal
conductance. On the other, the scattering formalism, in the limit
$d_{s}/\xi \rightarrow \infty$ and
${\cal T} \rightarrow 1$ gives that both ${\cal T}$ (${\cal T}'$)
and ${\cal T}_{\rm A}$ (${\cal T}_{\rm A}'$) vanish,
and so does ${\cal R}$ (${\cal R}'$), while ${\cal R}_{\rm A}$
(${\cal R}_{\rm A}'$) tends to unity
[see Eqs. (\ref{LOP}) and (\ref{IMP})].
As a result, $g_{LL}=g_{RR}=4e^{2}/h$ and $g_{LR}=g_{RL}=0$, and
Eq.
(\ref{ISOF}) becomes exactly the classical addition of resistances,
producing
half the ideal quantum conductance of the pure NS junction. This is due to the
fixing of  the chemical potential on the S-section to conserve the current, as in
the classical treatment.

In summary, we have shown that  the presence of a superconducting
segment in an otherwise normal system reduces the overall conductance once
the barriers between the superconducting and the normal parts become high
enough. Thus, the superconducting segments may push the system towards the
localizes insulating state.

From the appearance of the plots presented in Figs. \ref{EQ} and \ref{NEQ},
one
may be tempted to say that the system experiences a metal-insulator quantum
phase transition  from a finite to a vanishing conductivity at large
$d_{s}/ \xi$, when ${\cal T}$ decreases (which can be inferred to as 
``disorder increase"). We refrain
here from making such a statement and defer the discussion 
of such a quantum phase transition in the thermodynamic limit for the composite NS system to
future work.

\begin{acknowledgments}
We thank M. Schechter for many illuminating discussions. This work was
supported by the German Federal Ministry of Education and Research (BMBF)
within the framework of the German-Israeli project cooperation (DIP), and by
the Israel Science Foundation (ISF) and by the Converging Technologies Program
of the Israel Science Foundation (ISF), grant No 1783/07.
\end{acknowledgments}


\begin{thebibliography}{999}
\bibitem{COM1}


Here we use implicitly the Andreev approximation \cite{BTK} in
which the magnitudes of  the group velocities of all
quasiparticles equal the Fermi velocity.
\end{thebibliography}
\end{document}